\documentclass[aip,rsi,reprint,graphicx, floatfix]{revtex4-1}  

\usepackage{graphicx}
\usepackage{amsmath,amsthm,amssymb}
\usepackage{color}
\usepackage{amsfonts}
\usepackage{subfigure}
\usepackage{bm}
\usepackage{cleveref}

\DeclareRobustCommand{\uvec}[1]{{%
\ifcsname uvec#1\endcsname
\csname uvec#1\endcsname
\else
\bm{\hat{\mathbf{#1}}}%
\fi
}}

\newcommand{\imi}{\mathrm{i}}
\newcommand{\eue}{\mathrm{e}}
\newcommand{\bvec}[1]{\vec{\mathbf{#1}}}
\newcommand{\diff}{\mathop{}\!\mathrm{d}}

\begin{document}

\title{Noise characterization for resonantly-enhanced polarimetric vacuum magnetic-birefringence experiments} 

\author{M. T. Hartman}
\email[Michael T. Hartman: ]{michael.hartman@lncmi.cnrs.fr}
\affiliation{Laboratoire National des Champs Magn\'etiques Intenses (UPR 3228, CNRS-UPS-UGA-INSA), F-31400 Toulouse Cedex, France}
\author{A. Riv\`ere}
\affiliation{Laboratoire National des Champs Magn\'etiques Intenses (UPR 3228, CNRS-UPS-UGA-INSA), F-31400 Toulouse Cedex, France}
\author{R. Battesti}
\affiliation{Laboratoire National des Champs Magn\'etiques Intenses (UPR 3228, CNRS-UPS-UGA-INSA), F-31400 Toulouse Cedex, France}
\author{C. Rizzo}
\affiliation{Laboratoire National des Champs Magn\'etiques Intenses (UPR 3228, CNRS-UPS-UGA-INSA), F-31400 Toulouse Cedex, France}

\date{07 June 2017}
\revised{03 December 2017}

\begin{abstract}
In this work we present data characterizing the sensitivity of the Bir\'{e}fringence Magnetique du Vide (BMV) instrument. BMV is an experiment attempting to measure vacuum magnetic birefringence (VMB) via the measurement of an ellipticity induced in a linearly polarized laser field propagating through a birefringent region of vacuum in the presence of an external magnetic field.  Correlated measurements of laser noise alongside the measurement in the main detection channel allow us to separate measured sensing noise from the inherent birefringence noise of the apparatus.  To this end we model different sources of sensing noise for cavity-enhanced polarimetry experiments, such as BMV.  Our goal is to determine the main sources of noise, clarifying the limiting factors of such an apparatus. We find our noise models are compatible with the measured sensitivity of BMV. In this context we compare the phase sensitivity of separate-arm interferometers to that of a polarimetry apparatus for the discussion of current and future VMB measurements.
\end{abstract}


\maketitle 

\section{Introduction}

The search for effects of electromagnetic fields on light dates back to the beginnings of modern physics. In matter as a medium for light propagation, the Faraday effect~\cite{Faraday1846} has been known since the middle XIX century. At the turn of the century the possibility that, even in a vacuum, light interacts with an electromagnetic field has been taken into consideration, giving rise to a research field which is still open~\cite{Battesti2013}. The first motivation to look for such a phenomenon was the search for a magnetic moment of the photon.  It was eventually seen that an entire class of vacuum nonlinear optical effects are allowed within the framework of Quantum ElectroDynamics (QED)~\cite{Battesti2013}, and, in more general terms, other Non-Linear ElectroDynamics (NLED) theoretical frameworks~\cite{Fouche2016}.

Since its inception, the Michelson-Morley interferometer has attracted the attention of experimentalists seeking to measure a differential light velocity in the presence of a magnetic field. A first experiment has been performed by Morley himself around 1898 and others followed in the first half of the XX century~\cite{Battesti2013}. In recent years this topology has been refined up to the observation of gravitational waves on Earth~\cite{Abbott2016} reaching unprecedented sensitivities in interferometry.  Experimental proposals have been put forward in recent~\cite{Grote2015} and less recent years (see~[\cite{Battesti2013}] and references therein) hoping to take advantage of the technological progress in separate-arm interferometry in the domain of light and magnetic field interactions in vacuum.

In the 1970s the expected values for the index of refraction of light polarized parallel, $n_\parallel$, and perpendicular, $n_\perp$, to an applied external magnetic field were calculated~\cite{Bialynicka-Birula1970}, thanks to the previous works of Euler, Kochel, and Heisenberg~\cite{Euler1935,Heisenberg1936}. Following~\cite{Battesti2013}, one can write
\begin{align}
n_{\parallel} &= 1 + {c_{0,2}}\frac{B_{0}^2}{\mu_{0}}\\
n_{\perp} &=  1 + {4c_{2,0}}\frac{B_{0}^2}{\mu_{0}}
\end{align}
where the value of the lowest order coefficients of the development of the Heisenberg-Euler Lagrangian, $c_{2,0}$ and $c_{0,2}$, can be written as~\cite{Euler1935,Heisenberg1936}:
\begin{equation}
	c_{2,0} = \frac{2\alpha^2 \hbar^3}{45 m_{e}^4 c^5},\quad c_{0,2} = 7 c_{2,0}
\end{equation}
The QED predicted change in vacuum index of refraction, both parallel and perpendicular to an applied external field $B_0 [\mathrm{T}]$, can be thus be written:
\begin{align}
	\delta{n_{\parallel}} &= 1-n_{\parallel} \approx 9\times 10^{-24}\frac{B_{0}^2}{\mathrm{T}^2}\label{eqn:deltanpl}\\
	\delta{n_{\perp}} &= 1-n_{\perp} \approx 5\times 10^{-24}\frac{B_{0}^2}{\mathrm{T}^2}\label{eqn:deltanpr}
\end{align}

Following \eqref{eqn:deltanpl} and \eqref{eqn:deltanpr}, $\Delta n$ can be written
\begin{equation}
\Delta n = n_{\parallel} - n_{\perp} \approx 4\times 10^{-24}\frac{{B_0}^2}{\mathrm{T}^2}.
\label{Deltan}
\end{equation}

This form is analogous to one corresponding to the Cotton-Mouton effect, the linear magnetic birefringence in a medium discovered at the beginning of the XX century and studied in detail by A. Cotton and H. Mouton~\cite{Rizzo1997}.  Traditionally this measurement of $\Delta n$ is obtained via a measurement of the ellipticity, $\psi$, acquired by a linearly polarized laser beam of wavelength $\lambda$ propagating through the region, $L_B$, of magnetic field, $B$.  The resulting ellipticity is due to the phase shift between the two orthogonal polarization components of the light field,
\begin{equation}
	\gamma = \phi_{x} - \phi_{y},
\end{equation}
where $\phi_{x}$ is the phase accumulated in the laser field component polarized parallel to the magnetic field, and $\phi_{y}$ is the perpendicular component.  For the measurement in vacuum, one can write the induced ellipticity
\begin{equation}
\psi = \pi\frac{L_B}{\lambda} k_\mathrm{CM} B^2,
\end{equation}
where, as predicted by QED,  $k_\mathrm{CM}~\approx~4\times10^{-24}\,\mathrm{T}^{-2}$ is the so-called Cotton-Mouton constant of vacuum~\cite{Battesti2013}.

In 1979 Iacopini and Zavattini proposed the use of ultra-precise polarimetry to measure the anisotropy of vacuum in the presence of an external magnetic field\cite{Zavattini1979}, putting forward the development of an instrument that is more sensitive than a separated-arm interferometer, as the measurement concerns ``the phase difference between two components of the same laser beam, and not the phase difference between two spatially separated beams''\cite{Zavattini1979}. In contrast to a differential-arm measurement, their proposal precludes the possibility of measuring $n_{\parallel}$ and $n_{\perp}$ separately, measuring only the difference, $\Delta{n}$.  Since then, all attempts to measure vacuum magnetic birefringence (VMB) have been based on this seminal paper~\cite{Battesti2013}, the community agreeing implicitly with their point of view.  Currently, the most advanced polarimetry experiment is the one of the PVLAS collaboration~\cite{DellaValle2016}.

In this work, we present data taken with our BMV apparatus~\cite{Cadene2014} from separate, but correlated, measurements of the noise coming from the laser amplitude fluctuations, cavity coupling noise, and cavity mirror birefringence noise. Using these data, we model the different sources of noise in cavity-enhanced polarimetry for the observation of VMB.  We compare these results to the phase sensitivity of separate-arm interferometers.  In section \ref{sec:seperatearm_interferometers} we present a conceptual scheme for a VMB measurement in a separate-arm Michelson interferometer. The section \ref{sec-resonantly_enhanced_birefringence} is an overview of key formulae concerning resonantly enhanced birefringence measurement using a Fabry-Perot cavity.  The following section \ref{sec:measurement_of_linear_magnetic_birefringence} describes the basic principles of polarimetric measurement of the linear magnetic birefringence with special attention to our BMV experiment~\cite{Cadene2014}.  In section \ref{sec:apparent_birefringence_noise} we model the different sources contributing to the sensing noise of cavity enhanced polarimeters.  The results showing the measured/modeled sensing noise alongside the measured total birefringence sensitivity is given in section \ref{sec:results}.  Finally sections \ref{sec:discussion} and \ref{sec:conclusions} present the measured cavity birefringence fluctuations and discussion of the possible sources of such a noise, ending in perspectives and final conclusions.

\section{Separate-arm interferometers}
\label{sec:seperatearm_interferometers}
\begin{figure}
  \begin{center}
  \includegraphics[width=0.74\columnwidth]{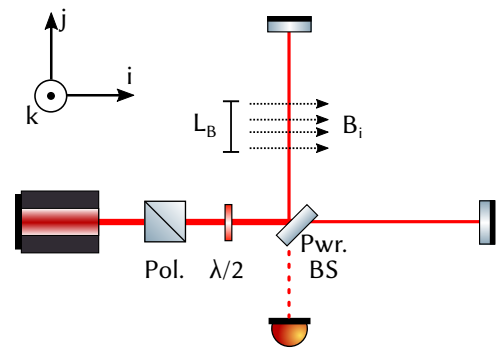}
  \end{center}
  \vspace{-9pt}
  \caption{Sketch of a conceptual Michelson-Morley interferometer setup to measure a variation of light velocity in vacuum in the presence of an external magnetic field $\mathbf{B}_\mathrm{i}$.}
  \vspace{-3pt}
  \label{fig:Michelson_Bfield}
\end{figure}

In figure \ref{fig:Michelson_Bfield}  we show a conceptual illustration of a Michelson-Morley interferometer setup to detect a variation in the velocity of linearly polarized light propagating through birefringnent vacuum in the presence of a transverse magnetic field, $\bvec{B}_\mathrm{i}$.  A modulation of the differential phase between the recombined beams produces a measurable interference pattern corresponding to the difference of light travel time through the two vacuum media in the differential arms.  This path-length modulation can be achieved through modulation of the light polarization angle, modulation of the magnetic field orientation, or by modulating the amplitude of the magnetic field.  The resulting phase difference can be written:
\begin{equation}
	\Delta{\phi} = \phi_\mathrm{i}-\phi_\mathrm{j} = \frac{2\pi}{\lambda}2L_B\delta n,
\label{PS-MMVMB}
\end{equation}
where $\delta n$ is the difference between the vacuum refractive indices of the two arms.

As a point of reference, the leading VMB polarimeter, PVLAS~\cite{DellaValle2016}, cites a sensitivity to change in index of refraction of $\tilde{n}\approx 3\times10^{-19}\,\frac{1}{\mathrm{\sqrt{Hz}}}$ around its detection frequency, $10\,\mathrm{Hz}$, using $1\,\mathrm{\mu}$ light propagating through a region of magnetic field $L_\mathrm{B}=1.6\,\mathrm{m}$.  One can back-calculate its sensitivity in terms of phase shift between the two polarization states of the laser field:
\begin{equation}
	\tilde{\phi}_\mathrm{PVLAS} =\frac{2\pi}{\lambda} L_\mathrm{B}\tilde{n} \approx 3\times10^{-12} \frac{\mathrm{rad}}{\mathrm{\sqrt{Hz}}}\:\quad\mathrm{at}\;\mathrm{10\,\mathrm{Hz}}
\end{equation}
For comparison, the field of gravitational-wave detection has advanced the state of the art in precision differential-arm interferometry~\cite{ligo_instrument2015}.  The $L_0 = 4\,\mathrm{km}$ long advanced LIGO (aLIGO) interferometers measure a strain noise of $\tilde{h} = \frac{\tilde{L}}{L_0}\approx8\times10^{-24}\,\frac{1}{\mathrm{\sqrt{Hz}}}$ at $\mathrm{200\,\mathrm{Hz}}$ (the center of their measurement band) increasing to $2\times10^{-22}\,\frac{1}{\mathrm{\sqrt{Hz}}}$ around $20\,\mathrm{Hz}$, the edge of their detection band\cite{Abbott2016}.  For direct comparison,  we write this in terms of sensitivity to phase delay between its differential arms:
\begin{align}
	\tilde{\phi}_\mathrm{LIGO} =\frac{2\pi}{\lambda} L_\mathrm{0}\tilde{h} &\approx 5\times10^{-12} \frac{\mathrm{rad}}{\mathrm{\sqrt{Hz}}}\:\quad\mathrm{at}\;20\,\mathrm{Hz}\\
	&\approx 2\times10^{-13} \frac{\mathrm{rad}}{\mathrm{\sqrt{Hz}}}\:\quad\mathrm{at}\;200\,\mathrm{Hz}
\end{align}
with LIGO using the same $\lambda\approx 1\,\mathrm{\mu}$ light as PVLAS and BMV.  This sensitivity is further discussed in comparison to the BMV appartus in section \ref{sec:comparison_of_bmv_and_ligo_phase_sensitivity}.

\section{Resonantly enhanced birefringence: key formulae}
\label{sec-resonantly_enhanced_birefringence}

Polarimeter VMB searches measure the differential phase between the two polarization states of light passing through a vacuum region in the presence of a magnetic field.  Current VMB experiments utilize a two-mirror resonant optical cavity to enhance the sensitivity to the differential phase shift resulting from vacuum birefringence.  The transfer function for an optical cavity can be expressed in terms of the laser frequency $\omega = kc$, where $c$ is the speed of light. For a field $E_\mathrm{in} = E_{\mathrm{i}}\eue^{-{\imi}\omega t}$ incident on the cavity of length $L$, the intracavity field can be expressed in terms of the cavity mirrors' amplitude reflectance ($r_1$, $r_2$) and transmittance ($t_1$, $t_2$) 
\begin{equation}
E_\mathrm{cav} = \frac{t_{1}\eue^{-{\imi}\frac{\omega L}{c}}}{1-r_{1}r_{2}\eue^{-{\imi}\frac{2\omega L}{c}}}E_\mathrm{in} \label{eqn-fieldxferfunc_cav}
\end{equation}
It is apparent in \eqref{eqn-fieldxferfunc_cav} that resonances occur when the laser frequency, $f$, is at integer multiples of the Free Spectral Range $(\mathrm{FSR} = \frac{c}{2L})$:
\begin{equation}
	f = \frac{\omega}{2\pi} = N\frac{c}{2L} = N(\mathrm{FSR}) \label{eqn-cav_resonance_freq}
\end{equation}
The cavity finesse, $\mathcal{F}$, is determined by the optical losses in the cavity and is defined as the ratio of the FSR to the linewidth, or Full-Width at Half-Maximum $(\mathrm{FWHM})$, of the resonance peak.
\begin{equation}
	\mathcal{F} \triangleq \frac{\mathrm{FSR}}{\mathrm{FWHM}} \approx \frac{\pi \sqrt{r_1 r_2}}{1-r_1 r_2}\label{eqn-Finesse_definition}
\end{equation}

The gain, $g_\mathrm{cav}$, of laser field inside an optical resonator can be written:
\begin{align}
g_\mathrm{cav} 	&= \left|\frac{E_\mathrm{cav}}{E_\mathrm{in}}\right| = \sqrt{\frac{t_{1}^{2}}{1-r_{1}r_{2}\left(\eue^{{\imi}\phi}+\eue^{-{\imi}\phi}\right) + r_{1}^{2}r_{2}^{2} }} \nonumber\\
	&=    \frac{t_{1}}{\left(1-r_{1}r_{2}\right)}   \sqrt{\frac{1}{1+\frac{4r_{1}r_{2}}{\left(1-r_{1}r_{2}\right)^2}\sin^2{\left(\frac{\phi}{2}\right)}} } \label{eqn-mag_trans_field}
\end{align}
where $\phi=2kL= \frac{4\pi}{\lambda}L = 2\pi\frac{f}{\mathrm{FSR}}$ is the cavity round-trip accumulated phase.  The frequency response of the amplitude can be seen more clearly to have a filtering effect by simplifying the second term in \eqref{eqn-mag_trans_field}, which we will call $\mathrm{F}_\mathrm{cav}$.   Near resonance one can write
\begin{equation}
\mathrm{F}_\mathrm{cav}(f) \approx \frac{1}{\sqrt{1+\frac{4r_{1}r_{2}}{\left(1-r_{1}r_{2}\right)^2}\frac{\pi^2}{\mathrm{FSR}^2}f^2}}= \frac{1}{   \sqrt{1+\left(\frac{f}{f_\mathrm{c}}\right)^{2}}   }
\end{equation}
where $f_\mathrm{c} = \frac{\mathrm{FWHM}}{2}$ is the cavity pole frequency.

The phase accumulation $(\Phi)$ of the intracavity field near resonance is given by the expression
\begin{equation}
	\Phi = \arctan\left(\frac{\Im{E_\mathrm{cav}}}{\Re{E_\mathrm{cav}}}\right) \label{eqn-Phi_cav}
\end{equation}
We can evaluate this by expanding cavity field \eqref{eqn-fieldxferfunc_cav}:
\begin{align}
	\frac{E_\mathrm{cav}}{E_\mathrm{in}} &= \frac{t_{1}\eue^{{\imi}\frac{\phi}{2}} - t_{1}r_{1}r_{2}\eue^{-{\imi}\frac{\phi}{2}}}{{1-r_{1}r_{2}\left(\eue^{{\imi}\phi}+\eue^{-{\imi}\phi}\right) + r_{1}^{2}r_{2}^{2} }} \nonumber \\
						&= \frac{t_{1}\cos{\frac{\phi}{2}}\left(1-r_{1}r_{2}\right)}{1-2r_{1}r_{2}\cos{\phi}+r_{1}^{2}r_{2}^{2}} + \imi\frac{t_{1}\sin{\frac{\phi}{2}}\left(1+r_{1}r_{2}\right)}{1-2r_{1}r_{2}\cos{\phi}+r_{1}^{2}r_{2}^{2}} \label{eqn-E_cav_re_im}
\end{align}
When the laser is near the cavity resonance, the approximate total phase accumulated can thus be written in terms of the rount-trip phase and the Finesse:
\begin{equation}
	\Phi \approx \frac{1+r_{1}r_{2}}{1-r_{1}r_{2}}\frac{\phi}{2} \approx \frac{\mathcal{F}}{\pi}\phi \\
\end{equation}

In a birefringent cavity the round-trip phase along the fast-axis, $\phi_{y}$, and slow-axis, $\phi_{x}$, for a laser on resonance leads to a total accumulated differential phase of
\begin{equation}
	\Gamma = \frac{\mathcal{F}}{\pi}\phi_{x} - \frac{\mathcal{F}}{\pi}\phi_{y} = \frac{\mathcal{F}}{\pi}\gamma \label{eqn-Gamma}
\end{equation}

\section{Measurement of linear magnetic birefringence}
\label{sec:measurement_of_linear_magnetic_birefringence}

\begin{figure}
  \begin{center}
  \includegraphics[width=\columnwidth]{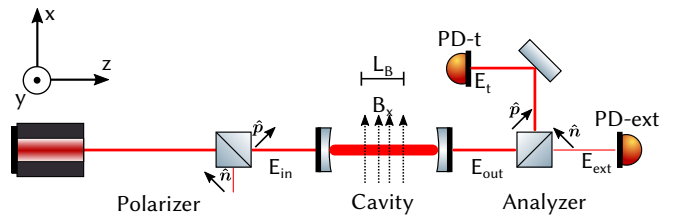}
  \end{center}
  \caption[Illustration of BMV]{An illustration of a cavity-enhanced VMB polarimeter. A laser is resonant in a Fabry-Perot cavity.  A magnetic field is applied transverse to the direction of laser propagation producing vacuum birefringence,  which is measured as the differential phase delay between the polarization components of the intracavity field.  The signal is enhanced by the phase response of the cavity.}
  \label{fig:VMB_Polarimeter_cavity_enhanced}
\end{figure}

The conceptual layout for a VMB search experiment is illustrated in figure \ref{fig:VMB_Polarimeter_cavity_enhanced}.  The setup uses a laser field propagating in the $\uvec{z}$ direction with amplitude $E_{\mathrm{i}}$ linearly polarized to some general angle $\theta$ with respect to the x-axis.  In the setup, the first polarizer is oriented to transmit light polarized in the direction $\uvec{p}$:
\begin{equation}
	\uvec{p} = \cos\theta\uvec{x}+\sin\theta\uvec{y}
\end{equation}
and to reject light polarized in the orthogonal direction, $\uvec{n}$:
\begin{equation}
	\uvec{n} = \sin\theta\uvec{x}-\cos\theta\uvec{y}
\end{equation}
In such a configuration the field entering the optical cavity can be described as
\begin{equation}
	\bm{E}_\mathrm{in} = E_{\mathrm{i}}\eue^{-\imi \omega t}\uvec{p}
\end{equation}

The laser frequency is actuated to remain resonant in a Fabry-Perot optical cavity.  A magnetic field is applied in the $\uvec{x}$ direction.

\begin{figure}
  \begin{center}
  \includegraphics[width=0.95\columnwidth]{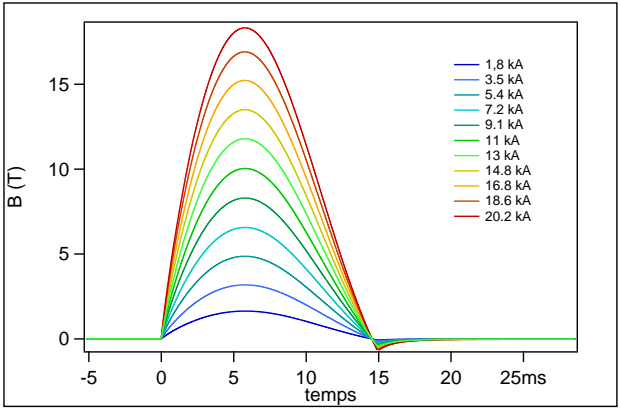}
  \end{center}
  \caption[The temporal profile of the XXL-coil pulsed field magnet to be used in the second generation BMV experiment.]{The temporal profile of the XXL-coil pulsed field magnet to be used in the second generation BMV experiment.  The temporal profile is selected to produce pulses below the cavity pole frequency, $f_\mathrm{c}$}
  \label{fig-xxl_temporal}
\end{figure}

To provide tangible parameter values we discuss the second generation BMV experiment, which will use a pulsed field magnet called the `XXL-coil'.  The temporal profile for the first test pulses of the XXL-coil can be seen in figure \ref{fig-xxl_temporal}, designed to maximize the interaction between the magnetic field and intracavity field.  These pulses delivered up to $18\,\mathrm{T}$ of field over an effective length of $L_B=0.319\,\mathrm{m}$.

In this setup, the intracavity laser field sees a round-trip differential phase retardation, $\gamma_\mathrm{v}$, between its polarization components lying along the fast optical axis $(\uvec{y})$ and slow optical axis $(\uvec{x})$ of the vacuum:
\begin{align}
	\gamma_\mathrm{v}	&= 2\frac{2\pi}{\lambda}L_\mathrm{B}k_\mathrm{CM} B^2 \label{eqn-gamma_vac}\\
						&\approx 5\times10^{-15}\,\mathrm{rad},\\
						&\quad(\lambda = 1\,\mathrm{\mu m},\: B^2 L_\mathrm{B}=100\,\mathrm{T}^2\mathrm{m})\nonumber
\end{align}	

Additionally, the cavity mirrors have an inherent birefringence~\cite{Bielsa2009}.  In each round trip the cavity receives an additional phase from from the end mirror $(\gamma_{\mathrm{m}2})$ as well as the input mirror $(\gamma_{\mathrm{m}1})$ giving a total differential phase per round trip of $\gamma_\mathrm{c} = \gamma_{\mathrm{m}2} + \gamma_{\mathrm{m}1}$.  The total round trip birefringence is thus
\begin{equation}
	\gamma = \gamma_\mathrm{c} + \gamma_\mathrm{v}.
\end{equation}
As we have recalled earlier, \eqref{eqn-Phi_cav}-\eqref{eqn-Gamma}, the many round trips in the cavity enhances this differential phase by a factor proportional to the finesse of the cavity:
\begin{equation}
	\Gamma = \frac{\mathcal{F}}{\pi}\gamma =\frac{\mathcal{F}}{\pi}(\gamma_\mathrm{c} + \gamma_\mathrm{v}), \label{eqn-Gamma_sum}
\end{equation}
with the net effect making the cavity appear as a waveplate of retardation $\Gamma$.

The signal is contained in the differential phase, $\Gamma$, between the two polarization states of the intracavity field,
\begin{equation}
	\bm{E}_\mathrm{cav} = E_{\mathrm{c}}\eue^{-\imi \omega t}\left(\eue^{\imi \Gamma}\cos\theta\uvec{x}+\sin\theta\uvec{y}\right).
\end{equation}
This phase shift is analyzed into into a measurable power change using a second polarizer called the `analyzer'.  The field incident on the analyzer is directly the field transmitted from the cavity, $t_2\bm{E}_\mathrm{cav}$. The analyzer is crossed with the input polarizer, oriented to transmit the $\uvec{n}$ component and to reflect the $\uvec{p}$ component.  For nominally amorphous mirrors $\Gamma$ can be very small, it is thus important to account for the imperfection of the analyzer, namely the analyzer's parallel and orthogonal polarization transmission and reflection coefficients: $t_{\parallel} \gg t_{\perp}$ and $r_{\perp} \gg r_{\parallel}$.

\subsection{The signal in transmission, $P_\mathrm{t}$}
\label{sec-the_reflected_signal}
The field reflected by the analyzer, $\bm{E}_\mathrm{t}$, is composed primarily of the light in the original polarization state, $\uvec{p}$. We can calculate this field as follows:
\begin{align}
	\bm{E}_\mathrm{t} 	&= r_{\perp}\left(t_2 \bm{E}_\mathrm{cav}\cdot\uvec{p}\right)\uvec{p} + r_{\parallel}\left(t_2 \bm{E}_\mathrm{cav}\cdot\uvec{n}\right)\uvec{n} \nonumber \\
						&= t_{2}E_\mathrm{c}\eue^{-\imi \omega t}\big[r_{\perp}\left(\eue^{\imi \Gamma}\cos^{2}\theta+\sin^{2}\theta\right)\uvec{p}\nonumber\\
						&\quad+ r_{\parallel}\left(\eue^{\imi \Gamma}\cos\theta\sin\theta-\sin\theta\cos\theta\right)\uvec{n}\big]
\end{align}
With the field in our pocket, we can calculate the power reflected by the analyzer in terms of the intracavity power, $P_\mathrm{cav}$,
\begin{align}
	P_\mathrm{t} 	&= \bm{E}_\mathrm{t}\cdot\bm{E}_\mathrm{t}^{\ast} \nonumber \\
					&= T_{2}P_\mathrm{cav}\big[r_{\perp}^2(\eue^{\imi\Gamma}\cos^{2}\theta+\sin^{2}\theta) (\eue^{-\imi\Gamma}\cos^{2}\theta+\sin^{2}\theta)\uvec{p}\cdot\uvec{p} \nonumber\\
					&\quad+ r_{\parallel}^{2}\sin\theta\cos\theta(\eue^{\imi \Gamma}-1) (\eue^{-\imi\Gamma}-1)\uvec{n}\cdot\uvec{n}\big]\nonumber \\
					&= T_{2}P_\mathrm{cav}\bigg[R_{\perp}\left(1-\sin^2(2\theta)\sin^2\left(\frac{\Gamma}{2}\right)\right)\nonumber\\
					&\quad+ R_{\parallel}\sin^2(2\theta)\sin^2\left(\frac{\Gamma}{2}\right)\bigg]
\end{align}
Now we can approximate this expression for small $\Gamma$:
\begin{align}
	P_\mathrm{t}	&\approx T_{2}P_\mathrm{cav}\bigg[R_{\perp}\left(1-\sin^2(2\theta)\left(\frac{\Gamma^2}{4}\right)\right)\nonumber\\
					&\quad + R_{\parallel}\sin^2(2\theta)\left(\frac{\Gamma^2}{4}\right)\bigg]
	\label{eqn-Pt_small_Gamma}
\end{align}
To maximize the effect of vacuum birefringence, we orient the polarisation to be at an angle $\theta=\frac{\pi}{4}$ with respect to the external field.
In this case \eqref{eqn-Pt_small_Gamma} becomes
\begin{align}
	P_\mathrm{t} 	&\approx T_{2}P_\mathrm{cav}\left[R_{\perp}\left(1-\frac{\Gamma^2}{4}\right) + R_{\parallel}\frac{\Gamma^2}{4}\right]\nonumber\\
			&\approx T_{2}P_\mathrm{cav}\left[R_{\perp}\left(1-\frac{\Gamma^2}{4}\right)\right], \label{eqn-Pt_small_Gamma_45deg}
\end{align}
noting that $R_{\parallel}\ll 1$ and $\Gamma^2 \ll 1$.
Here we see that $\frac{\Gamma^2}{4}$ is the fractional power moved from the $\uvec{p}$ to the $\uvec{n}$ polarization state.  We see this as a decrease in the $R_{\perp}$ term.  This is the total power in the other polarization state, but is suppressed by the small $R_{\parallel}$ of the analyzer.

\subsection{The signal in extinction, $P_\mathrm{ext}$}

The signal transmitted through the polarizer is comprised of the light that has been transformed into the $\uvec{n}$ polarization state due to the intracavity birefringence, as well as unwanted light in the $\uvec{p}$ polarization state that leaks through the polarizer.  The field in this extinction channel can be described as
\begin{align}
	\bm{E}_\mathrm{ext}	&= t_{\perp}(t_2 \bm{E}_\mathrm{cav}\cdot\uvec{p})\uvec{p} + t_{\parallel}(t_2 \bm{E}_\mathrm{cav}\cdot\uvec{n})\uvec{n} \nonumber \\
							&= t_{2}E_\mathrm{c}\eue^{-\imi \omega t}\big[t_{\perp}(\eue^{\imi \Gamma}\cos^{2}\theta+\sin^{2}\theta)\uvec{p} \nonumber\\
&+ t_{\parallel}(\eue^{\imi \Gamma}\cos\theta\sin\theta-\sin\theta\cos\theta)\uvec{n}\big]
\end{align}
We can now write down the power transmitted through the analyzer in terms of the intracavity power, $P_\mathrm{cav}$:
\begin{align}
	P_\mathrm{ext}	&= \bm{E}_\mathrm{ext}\cdot\bm{E}_\mathrm{ext}^{\ast} \nonumber \\
					&= T_2 P_\mathrm{cav}\bigg[T_{\perp}\left(1-\sin^2(2\theta)\sin^2\left(\frac{\Gamma}{2}\right)\right) \nonumber\\
					&+ T_{\parallel}\sin^2(2\theta)\sin^2\left(\frac{\Gamma}{2}\right)\bigg]
\end{align}
Again, we will approximate this expression for small $\Gamma$; additionally, it is worth rewriting this expression in terms of a measurable quantity, the analyzer's extinction ratio, $\sigma^2 \triangleq \frac{T_{\perp}}{T_{\parallel}}$, thus
\begin{equation}
	P_\mathrm{ext} \approx T_{\parallel}T_{2}P_\mathrm{cav}\left[\sigma^2\left(1-\sin^2(2\theta)\frac{\Gamma^2}{4}\right) + \sin^2(2\theta)\frac{\Gamma^2}{4}\right] \label{eqn-Pext_small_Gamma}
\end{equation}
Again we make the polarisation to be at an angle $\theta = \frac{\pi}{4}$ with respect to the magnetic field. In this case \eqref{eqn-Pext_small_Gamma} becomes
\begin{align}
	P_\mathrm{ext} 	&\approx T_{\parallel}T_{2}P_\mathrm{cav}\left[\sigma^2\left(1-\frac{\Gamma^2}{4}\right) + \frac{\Gamma^2}{4}\right]\nonumber\\
			&\approx T_{\parallel}T_{2}P_\mathrm{cav}\left[\sigma^2 + \frac{\Gamma^2}{4}\right],
 \label{eqn-Pext_small_Gamma_45deg}
\end{align}
Here we note again that $\sigma^2$ and $\Gamma^2$ are both much less than $1$.  Similarly, $\frac{\Gamma^2}{4}$ is the fractional power moved from the $\uvec{p}$ to the $\uvec{n}$ polarization state.  Here $\sigma^2$ is the fraction of power leaking from the undesired polarization state $(\uvec{p})$ into the $P_\mathrm{ext}$ signal.

\section{Apparent Birefringence Noise}
\label{sec:apparent_birefringence_noise}

When trying to measure an effect as small as the vacuum Cotton-Mutton effect it is necessary to understand and characterize the competing noise sources.  In our case, the signal is measured as the change in power $(P_\mathrm{ext})$ incident on the extinction channel photodetector $(\mathrm{PD}_\mathrm{ext})$.  Power fluctuations from sources other than vacuum birefringence are indistinguishable from the power variation resulting from a vacuum-birefringence induced ellipticity.

\subsection{The signal as variation in laser power}

We start the noise investigation by first examining how a time-varying round-trip cavity birefringence, $\delta{\gamma}$, produces a time-varying power, $\delta{P_\mathrm{ext}}$, incident on photodetector $\mathrm{PD_{ext}}$. Once we see how the $\delta{\gamma}$ couples to $\delta{P_\mathrm{ext}}$ we can see how unwanted fluctuations in power, $\delta{P_\mathrm{noise}}$, masquerade as an apparent fluctuation in birefringence, $\delta\gamma_\mathrm{noise}$.

For the case of aligning the laser field's input polarization to $45^{\circ}$ with respect slow axis we can fully write the power incident on the photodetector $\mathrm{PD_{ext}}$ as a function of a dynamical input power, $P_\mathrm{in}$, viewed as amplitude sidebands filtered by the cavity filter, $\mathrm{F}_\mathrm{cav}$, and the round trip differential phase accumulation, $\gamma$, enhanced by the cavity finesse, $\mathcal{F}$:
\begin{equation}
	P_\mathrm{ext}= T_{\parallel}\left(\mathrm{F}_\mathrm{cav}\sigma^2+\mathrm{F}_\mathrm{cav}^{2}\frac{\mathcal{F}^2}{\pi^2}\frac{\gamma^2}{4}\right)T_{2}T_{1}\frac{\mathcal{F}^2}{\pi^2}P_\mathrm{in}. \label{eqn-Pext_of_Pin}
\end{equation}
Many noise sources, such as beam motion and laser frequency noise, interfere with the coupling of the laser field into the cavity mode.  Since the cavity power couples to the output power through the (stable) transmission constant of the end mirror, $T_{2}$, it is more practical to write the signal in terms of the cavity power:
\begin{equation}
	P_\mathrm{ext} = T_{\parallel}\left(\sigma^2+\mathrm{F}_\mathrm{cav}\frac{\mathcal{F}^2}{\pi^2}\frac{\gamma^2}{4}\right)T_{2}P_\mathrm{cav} \label{eqn-Pext_of_Pcav}
\end{equation}

Power fluctuations vary around a mean DC power on the detector, $\bar{P}_\mathrm{ext}$.  At DC the cavity filter is $\mathrm{F}_\mathrm{cav}=1$.  In terms of the average cavity power, $\bar{P}_\mathrm{cav}$, and the mean round-trip intracavity birefringence, $\gamma_{0}$:
\begin{align}
	\bar{P}_\mathrm{ext} &= P_\mathrm{ext}(\gamma = \gamma_{0}; \mathrm{F}_\mathrm{cav}=1; P_\mathrm{cav}=\bar{P}_\mathrm{cav})\nonumber\\
					 &=T_{\parallel}\left(\sigma^2+\frac{\mathcal{F}^2}{\pi^2}\frac{\gamma_{0}^2}{4}\right)T_{2}\bar{P}_\mathrm{cav} \label{eqn-Pext_ave}
\end{align}

Next, we examine the behavior around this mean by expanding the power \eqref{eqn-Pext_of_Pcav} as a function of $\gamma$ about the bias point $\gamma_{0}$:
\begin{equation}
	P_\mathrm{ext}(\gamma) = P_\mathrm{ext}(\gamma_{0})+\frac{1}{1!}\frac{\diff P_\mathrm{ext} }{\diff \gamma}\bigg|_{\substack{\gamma_{0}}}(\gamma-\gamma_{0}) + \mathcal{O}
\end{equation}
In the case of pulsed field magnets, we are interested in the dynamical birefringence.  We therefor want to rewrite this in terms of $\delta{P_\mathrm{ext}} = P_\mathrm{ext}(\gamma) -  P_\mathrm{ext}(\gamma_{0})$ and $\delta{\gamma} = \gamma-\gamma_{0}$.  Making the assumption the expected vacuum birefringence is much less than the static birefringence of the mirrors ($\delta{\gamma} \ll \gamma_{0}$) we can neglect $\mathcal{O}$ and this expression becomes
\begin{equation}
	\delta{P_\mathrm{ext}} = \frac{\diff P_\mathrm{ext} }{\diff \gamma}\bigg|_{\substack{\gamma_{0}}} \delta{\gamma} = T_{\parallel}\left(2 \mathrm{F}_\mathrm{cav}\frac{\mathcal{F}^2}{\pi^2}\frac{\gamma_{0}}{4}\right)T_{2}P_\mathrm{cav}\delta{\gamma} \label{eqn-P_expanded}
\end{equation}
We normalize this expression \eqref{eqn-P_expanded} by dividing by the average power \eqref{eqn-Pext_ave} to find the relative power change due to $\delta{\gamma}$:
\begin{equation}
	\frac{\delta{P_\mathrm{ext}}}{\bar{P}_\mathrm{ext}} = \mathrm{F}_\mathrm{cav}\frac{2\gamma_{0}}{\frac{4\pi^2}{\mathcal{F}^2}\sigma^2+\gamma_{0}^2}\delta{\gamma} \label{eqn-RPN_of_gamma}
\end{equation}
Conversely we can write the apparent birefringence fluctuations as function of relative power fluctuations.  Since we are concerned with the spectra of the signals, we switch to the convention of writing the linear spectral densities (LSD), $\delta{g}(f)$ as $\tilde{g}$
\begin{equation}	 \tilde{\gamma}=\frac{1}{\mathrm{F}_\mathrm{cav}}\frac{\frac{4\pi^2}{\mathcal{F}^2}\sigma^2+\gamma_{0}^2}{2\gamma_{0}}\frac{\tilde{P}_\mathrm{ext}}{\bar{P}_\mathrm{ext}}. \label{eqn-gamma_of_rpn}
\end{equation}

\subsection{Intracavity power noise}

Fluctuations in the intracavity laser field couple to apparent birefringence noise directly.  Here we will refer to unsuppressed intracavity power noise as residual (res) power noise.

We are interested in the power fluctuations measured at $\mathrm{PD}_\mathrm{ext}$, $\tilde{P}_\mathrm{ext:res}$, due to intracavity power fluctuations, $\tilde{P}_\mathrm{cav}$, alone.  At some static round-trip cavity birefringence, $\gamma_0$ we see
\begin{equation}	
\tilde{P}_\mathrm{ext:res}=T_{\parallel}\left(\sigma^2+\mathrm{F}_\mathrm{cav}\frac{\mathcal{F}^2}{\pi^2}\frac{\gamma_{0}^2}{4}\right)T_{2}\tilde{P}_\mathrm{cav}
\end{equation}
We can then compute the coupling of intracavity residual power noise to apparent birefringence noise using the relationship \eqref{eqn-gamma_of_rpn}, giving:
\begin{equation}	
\tilde{\gamma}_\mathrm{res}=\frac{1}{\mathrm{F}_\mathrm{cav}}\frac{2\pi^2\sigma^2}{\mathcal{F}^2\gamma_0}\frac{\tilde{P}_\mathrm{cav}}{\bar{P}_\mathrm{cav}} + \frac{\gamma_0}{2}\frac{\tilde{P}_\mathrm{cav}}{\bar{P}_\mathrm{cav}}\label{eqn:gamma_residual}
\end{equation}

We see from the two terms in \eqref{eqn:gamma_residual} that unsuppressed laser power noise couples into our signal through two mechanisms.  The first is leakage of the undesired polarization state $\uvec{p}$ through an imperfect extinction ratio,  $\sigma^2$, of the analyzer.  The second is the coupling of laser power into the signal's polarization state, $\uvec{n}$, via a static birefringence of the cavity, $\gamma_{0}$.

\subsection{NEP noise}
The photodetector has several inherent sources of detection noise.  Typical sources of electrical noise include dark-current shot noise, photo-current shot noise, Johnson noise of the resistor, and also (in the case of amplified detectors) amplifier electronic noise.  The photodetector manufacturer usually calculates and measures the sum of the detector's intrinsic noise sources and gives the value as the equivalent value if measured as optical laser power noise.  This value is called the photodetector's Noise Equivalent Power (NEP) and is in units of $[\frac{\mathrm{W}}{\mathrm{\sqrt{Hz}}}]$.

This relative power noise at $\mathrm{PD_{ext}}$ is simply the extinction photodetector's NEP, $\tilde{P}_\mathrm{NEP}$, divided by the average incident, $\bar{P}_\mathrm{ext}$, from which we can calculate the expected contribution to apparent birefringence noise:
\begin{equation}	
\tilde{\gamma}_\mathrm{NEP}=\frac{1}{\mathrm{F}_\mathrm{cav}}\frac{2\pi^2\tilde{P}_\mathrm{NEP}}{\gamma_0\mathcal{F}^2  T_\parallel T_2 \bar{P}_\mathrm{cav}}
\end{equation}

\subsection{Shot noise}

Shot noise, or photon counting noise, is the name given to the fluctuation in the rate of photons incident on a photon detector. The average laser power $\bar{P}\,[\mathrm{W}]$ is given by the product of the single photon energy, $\frac{hc}{\lambda}\,[\mathrm{J}]$, and the average rate of photons counted, $\bar{N}_{\gamma}\,[\frac{1}{\mathrm{s}}]$; or, the average rate of photons is measured by:
\begin{equation}
	\bar{N}_\mathrm{\gamma} = \frac{\bar{P}}{\frac{hc}{\lambda}}
\end{equation}

The variance of counting is determined by Poissonian statistics.  The standard deviation of this distribution is given by the square root of the average rate, $\sigma_{\gamma}=\sqrt{\bar{N}_{\gamma}}$.  A fluctuation in photon incidence rate is, at its core, a fluctuation in laser power.  The single-sided linear spectral density of power fluctuations is given by:
\begin{equation}
	\tilde{P}_\mathrm{shot} = \sqrt{2}\frac{hc}{\lambda}\sqrt{\bar{N}_{\gamma}} = \sqrt{2\frac{hc}{\lambda}\bar{P}}
\end{equation}
We can then compute the expected shot noise contribution to be
\begin{equation}	
\tilde{\gamma}_\mathrm{shot}=\frac{1}{\mathrm{F}_\mathrm{cav}}\frac{\pi}{\mathcal{F}} \sqrt{ \frac{2hc}{T_\parallel \lambda T_2 \bar{P}_\mathrm{cav}}\left( \frac{4\pi^2\sigma^2}{\mathcal{F}^2\gamma_{0}^{2}}+1 \right) }
\end{equation}

\subsection{Digitization noise}
Voltage noise inserted between the photodetector and analog-to-digital converter (ADC) appears directly as laser power noise, mitigated by the response of the detector.  At the data acquisition card the voltage recorded is proportional to the laser power incident on the photodetector, with the constants of proportionality being the photodiode's responsivity $R_\mathrm{PD}$, and the detector's transimpedance gain, $G_\mathrm{PD}$.  We can write the apparent laser power noise due to an ADC noise, $\tilde{V}_\mathrm{ADC}$:
\begin{equation}
	\tilde{\mathrm{P}}_\mathrm{ADC} = \frac{1}{G_\mathrm{PD}R_\mathrm{PD}}\tilde{V}_\mathrm{ADC}
\end{equation}

In an ideal ADC there is still an inherent noise to digitizing an analog signal.  This is the result of recording a continuous signal at specific instants separated by the sampling time $T_\mathrm{s} = \frac{1}{f_\mathrm{s}}$ and with an amplitude recorded at discrete `counts' at integer multiples of the least significant bit voltage,
\begin{equation}
	V_\mathrm{LSB} = \frac{\Delta{V}_\mathrm{ADC}}{2^{N}}, 
\end{equation}
for an ADC with an input voltage range of $\Delta{V}_\mathrm{ADC}$ and a resolution of $N$ effective bits. The LSD of the voltage noise from digitization is
\begin{equation}
	\tilde{V}_\mathrm{ADC} = \sqrt{\tilde{S}_\mathrm{ADC}} = \frac{V_\mathrm{LSB}}{\sqrt{6 f_\mathrm{s}}}	\label{eqn-LSD_digi_noise}
\end{equation}

We can thus write the apparent birefringence noise expected due to the signal quantization of the extinction channel photodetector with an ideal ADC as
\begin{equation}	
	\tilde{\gamma}_\mathrm{ADC}=\frac{1}{\mathrm{F}_\mathrm{cav}}\sqrt{\frac{2}{3 f_\mathrm{s}}}\frac{2^{-N}\pi^2\Delta{V}_\mathrm{ADC}}{G_\mathrm{PD}R_\mathrm{PD}\gamma_0\mathcal{F}^2  T_\parallel T_2 \bar{P}_\mathrm{cav}}
\end{equation}

\subsection{Laser frequency noise}

In a birefringent cavity, the resonance frequency depends on the polarization of the intracavity laser field~\cite{Berceau2012}. The difference between the resonance frequency for ordinary beam and the extraordinary beam, $\delta \nu$, is proportional to $\gamma_{0}$. In BMV, the laser frequency is locked to the resonance the ordinary beam, thus the extraordinary beam is slightly off resonance. The ratio between the intensity of extraordinary to ordinary light is~\cite{Berceau2012}
\begin{equation}
	a = \frac{1}{1+\frac{4\mathcal{F}^2}{\pi^2}\sin^2(\gamma_{0})}.
\end{equation}
Small errors in the coupling of the laser to the longitudinal mode of the cavity can produce amplitude modulation in the extraordinary resonant field.  If $\delta \nu_n$ is the cavity locking frequency noise, the $a$ ratio can be written as~\cite{Berceau2012}
\begin{equation}
	a = \frac{1+\frac{4\mathcal{F}^2}{\pi^2}\sin^2(\frac{\pi}{\mathcal{F}}\frac{\delta \nu_n}{\mathrm{FWHM}})}{1+\frac{4\mathcal{F}^2}{\pi^2}\sin^2(\gamma_{0}-\frac{\pi}{\mathcal{F}}\frac{\delta \nu_n}{\mathrm{FWHM}})}
\end{equation}
Typically, $\frac{\pi}{\mathcal{F}}\frac{\delta \nu_n}{\mathrm{FWHM}} <  \gamma_{0} \ll 1$ and therefore
\begin{equation}
	a \approx 1-\frac{4\mathcal{F}^2}{\pi^2}\left(\gamma_{0}^2-2\gamma_{0}\frac{\pi}{\mathcal{F}}\frac{\delta \nu_n}{\mathrm{FWHM}}\right).
\end{equation}
This longitudinal mode mismatch appears as a fluctuation in birefringence:
\begin{equation}
\gamma_\mathrm{freq} \approx \frac{\mathcal{F}}{\pi}\gamma_0^2\frac{\delta \nu_n}{\mathrm{FWHM}}
\end{equation}

Applying some typical experimental parameter values ($\frac{\mathcal{F}}{\pi}= 10^5,\:\gamma_0=10^{-8}\,\mathrm{rad},\:\delta{\nu} = 1\,\frac{\mathrm{mHz}}{\sqrt{\mathrm{Hz}}},$ and $\mathrm{FWHM} =100\,\mathrm{Hz}$~\cite{Berceau2012}\cite{Cantatore1995}), we expect this to produce birefringence noise on the order of $\gamma_\mathrm{freq}\approx 10^{-17}\frac{\mathrm{rad}}{\sqrt{\mathrm{Hz}}}$, $2$ orders of magnitude below the expected level of vacuum birefringence.

\section{Results}
\label{sec:results}

\begin{figure}
	\begin{center}
	\includegraphics[width=0.71\columnwidth]{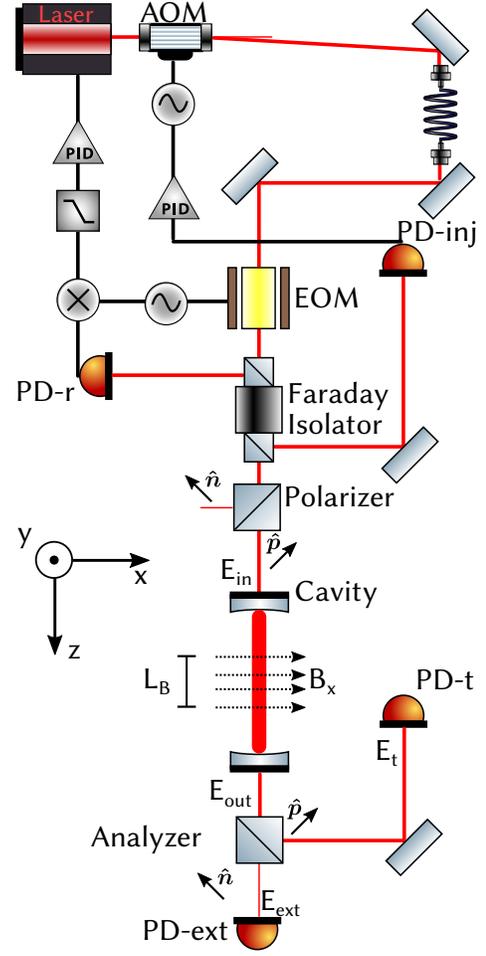}
	\end{center}
	\vspace{-6pt}
	\caption[Simplified illustration of the BMV experiment.]{Simplified illustration of the BMV experiment.}
	\label{fig:portrait_bmv_diagram_sans2ndpwrstab} 
\end{figure}

Here we compare the measured birefringence fluctuations in BMV to the correlated measurements and models of sensing noise outlined in this paper.  The layout of the BMV experiment is shown in figure \ref{fig:portrait_bmv_diagram_sans2ndpwrstab}.  The input optics start with a $0.5\,\mathrm{W},\:\lambda=1064\,\mathrm{nm}$ Nd:YAG laser.  The beam passes through an acousto-optic modulator (AOM) for power and frequency actuation before being sent through a single-mode fiber.  The resulting beam is phase-modulated at $\Omega_\mathrm{EOM}=10\,\mathrm{MHz}$ in an electro-optic modulator (EOM) to apply frequency sidebands.  The carrier and frequency sidebands are sent through a Faraday isolator, where a pickoff after the isolator sends light to photodetector PD-inj, used in-loop, feeding-back to the AOM for the power stabilization.

\begin{figure}
  \begin{center}
  \includegraphics[width=\columnwidth]{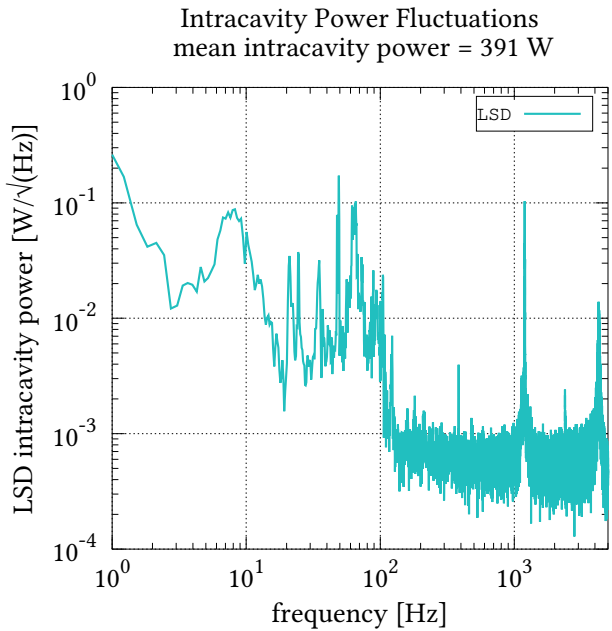}
  \end{center}
  \caption[Linear spectral density of the intracavity laser power.]{Linear spectral density of the intracavity laser power. }
  \label{fig:dsp_0003_intracavityLSD}
\end{figure}

The main polarimeter optics employ a $\sigma^2=2\times10^{-7}$ polarizer after which $\approx 20\,\mathrm{mW}$ of laser power is incident on the $\mathcal{F} = 440\:000$ cavity.  The cavity reflected light is diverted to the PD-r by the Faraday isolator where the photodetector's RF signal is demodulated with $\Omega_\mathrm{EOM}$ to produce the error signal in a feedback loop to actuate the laser frequency, keeping it tuned to the cavity's resonant frequency in an implementation of the Pound-Drever-Hall (PDH) technique~\cite{Black2001}.  The cavity transmitted light is passed through a matched polarizer (the `analyzer') to analyze polarization changes into laser power changes.  The transmission channel is detected at PD-t to measure common fluctuations in the intracavity field; the spectrum of intracavity power noise is plotted in figure \ref{fig:dsp_0003_intracavityLSD}.  Finally, the extinction channel is measured by PD-ext, monitoring changes in the polarization of the laser field.

\begin{figure}
  \begin{center}
  \includegraphics[width=\columnwidth]{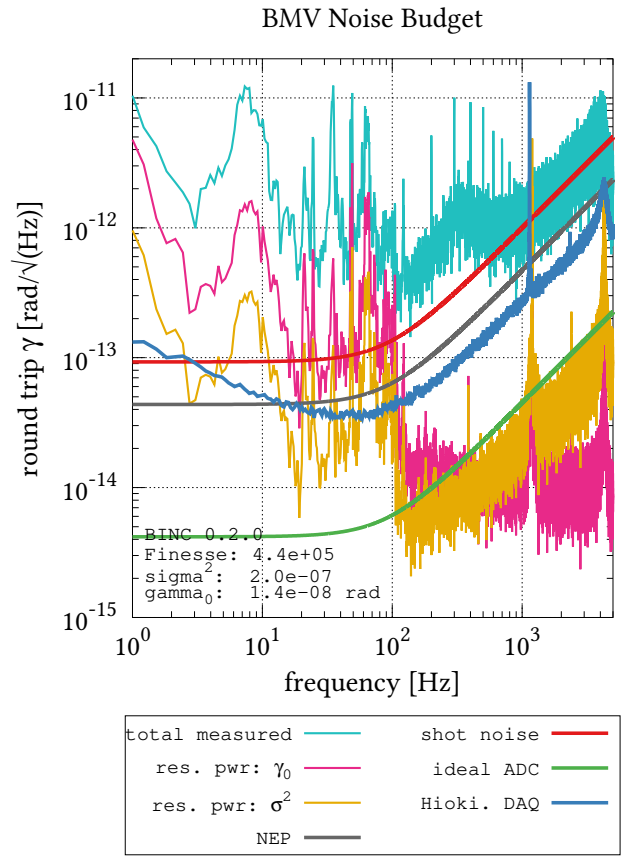}
  \end{center}
  \caption[The BMV noise budget.]{The BMV noise budget.  The measured sensitivity (cyan) is compared to the modeled noise sources, discussed in the text.}
  \label{fig:dsp_0003}
\end{figure}

We present in figure \ref{fig:dsp_0003} the measured birefringence spectrum along with the modeled sensing noise components. The linear spectral density of the measured intracavity birefringence fluctuations in radians per square root of hertz is plotted in {\em cyan}. In {\em magenta} is the measured noise contribution due to intracavity power noise coupling into the main signal polarization state through the cavity static birefringence, $\gamma_0$.  The coupling of intracavity power noise through by leakage through the analyzer's extinction ratio, $\sigma^2$, is plotted in {\em yellow}. These two noise models are derived from correlated measurements of the intracavity laser power (FIG. \ref{fig:dsp_0003_intracavityLSD}) and measurements of the respective optical parameters.  Noise due to the extinction channel photodetector electronic noise (NEP) is plotted in {\em grey}; BMV uses the low noise Femto OE-200-IN1 photoreceiver with an NEP $\approx 9\,\frac{\mathrm{fW}}{\sqrt{\mathrm{Hz}}}$. The contribution from shot noise is plotted in {\em red}.   The spectrum of voltage noise on the Hioki digital oscilliscope used for data acquisition (DAQ) was taken by measuring the input shorted with a $50\,\mathrm{ohm}$ terminator; the equivalent birefringence spectral noise density due to this DAQ noise is plotted in {\em blue}.  The specification for the Hioki DAQ gives the voltage resolution as $V_\mathrm{LSB} = \frac{\Delta{V}_\mathrm{ADC}}{1600}$, compatible with the measured noise.  This equates to roughly $N=10.6$ effective bits.  For a point of reference, the calculated noise contribution from an ideal 14-bit ADC is plotted in {\em green}.

The shown data set was taken after rotating the cavity mirrors to align the optical axis of the cavity to the polarization of the laser field, minimizing the ellipticity induced by the cavity.  The mean intracavity round-trip birefringence was measured to be $\gamma_0 = 1.4\times10^{-8}\,\mathrm{rad}$. This effort is rewarded with a significantly reduced contribution from residual laser power noise coupling through the $\gamma_0$ channel, albeit this remains the largest modeled noise term at low frequencies (below $100\,\mathrm{Hz}$).  With a low light level on the extinction detector, PD-ext, shot noise is a large contribution to the total sensing noise, and becomes the limiting noise source at high frequencies.

\section{Discussion}
\label{sec:discussion}

In this paper we modeled and measured various sources of sensing noise and showed how they appear as birefringence noise, limiting the sensitivity of cavity enhanced polarimeters. From this we can differentiate actual cavity birefringence fluctuations from sensing noise.  Here we discuss possible causes of dynamical cavity birefringence and compare the phase sensitivity of polarimeters to state of the art separate-arm interferometers, namely the LIGO gravitational-wave detector, in the context of measuring vacuum birefringence.

\subsection{Dynamical cavity birefringence}

\begin{figure}
  \begin{center}
  \includegraphics[width=\columnwidth]{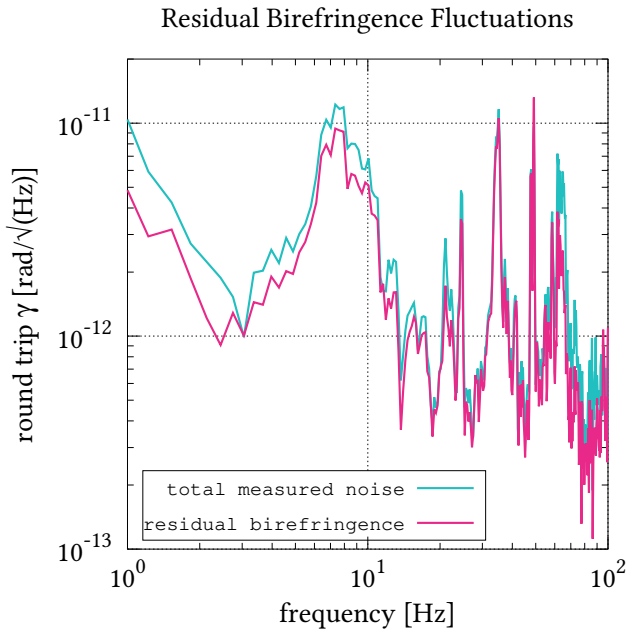}
  \end{center}
  \caption[Measured residual birefringence in the BMV polarimeter.]{Measured residual birefringence in the BMV polarimeter.}
  \label{fig:dsp_0003_residual}
\end{figure}
As was detailed in the previous sections, the measured sensitivity can be modeled only partially by the sensing noise.  Figure \ref{fig:dsp_0003_residual} shows actual dynamical birefringence of cavity as the difference between the measured birefringence and the correlated measurements of sensing noise. This residual noise is calculated by subtracting the time series of the auxiliary measured channels (scaled by the appropriate measured optical parameters: $\sigma^2$, $\gamma_0$, and $\mathcal{F}$ as outlined section \ref{sec:apparent_birefringence_noise}) from the time series of the measured birefringence.  The plot shows the linear spectral density of the resulting time series is at the level of $10^{-12}\,\frac{\mathrm{rad}}{\sqrt{\mathrm{Hz}}}$ for a large part of the frequency spectrum. This source of cavity birefringence noise is of import to cavity enhanced polarimetry in general.  The PVLAS experiment proposes to check temperature effects on cavity mirror birefringence~\cite{DellaValle2016} and has already measured for the effect of intracavity beam motion in their experiment, finding it not to be their dominating noise source ~\cite{DellaValle2016}.   As it has a different geometry and measurement frequency, these potential noise sources must be characterized in the BMV apparatus as well.  We discuss them here, briefly.

A potential source of dynamical birefringence in a Fabry-Perot cavity is the motion of the laser beam on the surface of the cavity mirrors.  Dielectric mirrors have been shown to have an inherent anisotropy~\cite{Bielsa2009}. The phase shift induced on an incident polarized laser beam may depend on the point of the mirror surface reflecting the light. Thus, if the pointing of the laser beam fluctuates with respect to the mirror position, a fluctuation in birefringence could be observed.  While recently studies of the mirror birefringence's positional dependence have been limited, a two-dimensional surface characterization of this effect has been reported~\cite{Micossi1993}. As, to the best of our knowledge, this is the only characterization of this kind we use it as a guide to estimate an order of magnitude of birefringence noise.  The referenced chart indicates that a translation of $1\,\mathrm{mm}$ results in about a $1\%$ variation in $\gamma$. Following ref.~\cite{Bielsa2009}, high finesse dielectric mirrors have shown a birefringence per reflection on the order of $10^{-6}\,\mathrm{rad}$. Assuming a linear response, a translation noise of about $100\,\frac{\mathrm{nm}}{\sqrt{\mathrm{Hz}}}$ would be needed to reproduce our measured cavity birefringence noise, a value compatible with the mechanical motion measured in commercial kinematic mirror mounts (see e.g.~\cite{Hartman2014}). Furthermore, fluctuations in the incidence angle must be taken into consideration, as a non-zero incidence angle produces a differential phase shift between the polarization states that is proportional to the square of the angle of incidence~\cite{Bouchiat1982}.To evaluate the noise induced by fluctuations in incident angle, $\theta$, let us recall that the corresponding birefringence per reflection can be written as $\gamma_\theta = \gamma_{i}\theta^2$~[\cite{Carusotto1989}]. Again, to induce the noise level that we observe, an angle of about $10^{-3}\,\frac{\mathrm{rad}}{\sqrt{\mathrm{Hz}}}$ would be necessary. In a $2\,\mathrm{m}$ cavity this would translate to a linear shift of $2\,\frac{\mathrm{mm}}{\sqrt{\mathrm{Hz}}}$, which would need to be much larger than translational estimate above.

Thermal effects on the reflective surface of the cavity mirrors is another potential source of intracavity mirror birefringence noise. Photon absorption produces localized temperature changes on the mirror surface, changing the optical path length experienced as the laser passes through the outer coating layers. The phase retardation between the laser field's polarization states should change proportionally, but this phenomenon has yet to be measured. Variations in power incident on the cavity mirrors translate into temperature noise and potentially into birefringence noise. One can naively estimate this birefringence variation by assuming that $\frac{\delta\gamma_\mathrm{i}}{\gamma_i} \approx \frac{\delta L}{L_0}$, where $L_0$ represents the mirror layer thickness. $\delta{L}$ can be written as $\alpha \delta{T} L_0$ where $\delta{T}$ is the variation of the mirror temperature and $\alpha$ the thermal expansion coefficient. Again, to reproduce the observed birefringence noise, where $L_0 \approx \lambda \approx 1\,\mathrm{\mu}$, $\gamma_i \approx 10^{-6}\,\mathrm{rad}$, and $\alpha \approx 10^{-6}\,\mathrm{K}^{-1}$~[\cite{Evans2008}].  To explain our measured intracavity birefringence (fig. \ref{fig:dsp_0003_residual})  one would need a $\delta T$ of the order of $1\,\frac{\mathrm{K}}{\sqrt{\mathrm{Hz}}}$, an unrealistically high value in BMV's measurement range of $10-100\,\mathrm{Hz}$; however, this potentially describes future fundamental noise floor resulting from the spectrum of heating due to intracavity shot noise, and warrants future investigation.

\subsection{Comparison of BMV and LIGO phase sensitivity}
\label{sec:comparison_of_bmv_and_ligo_phase_sensitivity}

In contrast to differential arm interferometry (such as a Michelson interferometer), a cavity-enhanced polarimetry measurement of vacuum birefringence has the advantage of being, to first order, insensitive to interferometric path length changes.  As was outlined in the introduction, the polarimeter topology is restricted to the measurement of the difference between the indicies of refraction perpendicular and parallel to a transverse magnetic field.  This limits the measurable physics in testing non-linear electrodynamic theories \cite{Fouche2016}.  Furthermore, as this paper has shown, the sensitvity of polarimeter searches are subject to fluctuations in birefringence of the interferometer mirrors.  As such, the measurement of vacuum polarization in a split-arm interferometer has interest.

Large-scale ultra precise resonantly enhanced-Michelson interferometers, such as the LIGO interferometers\cite{ligo_instrument2015}, have been built for the observation of gravitational waves.  For this purpose they seek to have the highest strain $(h=\frac{\delta L}{L_0})$ sensitivity to differential length changes, and have been designed with long baseline arm lengths $L_0=4\,\mathrm{km}$.  One can imagine building a smaller scale interferometer more suited to the purpose of measuring magnetic field excited vacuum polarization along its baseline arm.  However, gravitational-wave detectors make a good point of comparison for sensitivity to VMB measurements, being the most sensitive differential interferometers ever built, particularly in their most sensitive frequency band, $30\,\mathrm{Hz}-3\,\mathrm{kHz}$, which is compatible with characteristic frequencies of pulsed magnets, $30-200\,\mathrm{Hz}$.

\begin{figure}
  \begin{center}
  \includegraphics[width=\columnwidth]{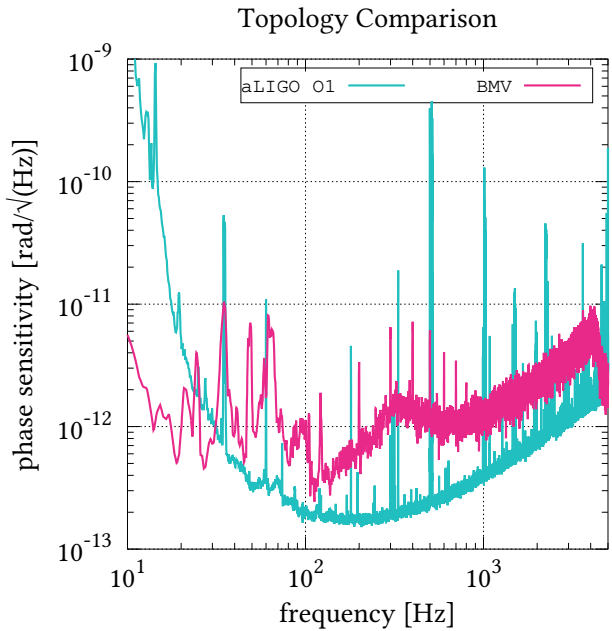}
  \end{center}
  \caption[Comparison of phase sensitivities for current experiments in Michelson (aLIGO) and polarimeter (BMV) topologies.]{Comparison of phase sensitivities for current experiments in Michelson (aLIGO)\cite{Abbott2016, ligo_strain} and polarimeter (BMV) topologies.}
  \label{fig:dsp_0003_LIGO_BMV_medres}
\end{figure}

To compare the instruments' sensitivities we state the aLIGO strain sensitivity, $\tilde{h}$, in terms of sensitivity to differential arm intracavity phase delay, $\widetilde{\delta{\phi}}$:
\begin{equation}
	\widetilde{\delta{\phi}} = \frac{2\pi}{\lambda} L_0\tilde{h}
\end{equation}
where aLIGO uses the same $\lambda=1064\,\mathrm{nm}$ laser light as BMV.  Using the strain sensitivity representative of the aLIGO Livingston observatory in 2015 \cite{Abbott2016, ligo_strain}, we plot the equivalent aLIGO phase sensitivity alongside the measured phase sensitivity of the BMV apparatus in figure \ref{fig:dsp_0003_LIGO_BMV_medres}.  

With extraordinary investment and engineering \cite{ligo_instrument2015}, Michelson interferometers can exceed the current phase sensitivity achieved by the BMV polarimeter, an interesting notion for the future study of non-linear electrodynamics.  In either case, the QED predicted additional accumulated phase perpendicular and parallel to the external magnetic field for $1\,\mu$ light through a region of $B^2 L = 100\,{\mathrm{T}^2\mathrm{m}}$ are
\begin{align}
	\phi_\parallel &= \frac{2\pi}{\lambda}\frac{14\alpha^{2}\hbar^{3}}{45m_\mathrm{e}^{4}c^{5}}\frac{B^{2}}{\mu_0}L_B\approx5.5\times10^{-15}\,\mathrm{rad}\\
	\phi_\perp &= \frac{2\pi}{\lambda}\frac{8\alpha^{2}\hbar^{3}}{45m_\mathrm{e}^{4}c^{5}}\frac{B^{2}}{\mu_0}L_B\approx3.1\times10^{-15}\,\mathrm{rad}
\end{align}
clearly calling for future improvements in both optical sensitivity and pulsed magnet technique.  We note that the pulsed magnet signal has a predictable form, and that the signal to noise ratio is improved by making a correlated average of data taken over many pulses \cite{Cadene2014}.

\section{Conclusions}
\label{sec:conclusions}

We have modeled several noise sources for cavity-enhanced precision polarimetry vacuum birefringence searches.  We found our sensing-noise budget to be compatible with the measured sensitivity of the BMV polarimeter.  It offers an explanation of the measured sensitivity at high frequencies, however, it appears that the sensing noise cannot fully explain the measured birefringence.  This suggests that, at low frequencies, BMV's sensitivity is limited by cavity mirror birefringence fluctuations. This noise (figure \ref{fig:dsp_0003_residual}) has a similar spectrum and is correlated in time with measured intracavity power noise shown (figure \ref{fig:dsp_0003}) suggesting that both effects could have common source, such as relative motion between the cavity laser mode and position of the cavity mirrors.

Beam position and thermal effects are sources of birefringence noise which have not yet been studied experimentally. This is a required work to better understand and subsequently mitigate their impact on the sensitivity of precision polarimetry experiments. Further more, one could circumvent such a noise source if the intrinsic birefringence of dielectric mirrors could be reduced during mirror fabrication. Unfortunately, the origin of birefringence in dielectric mirrors is still unexplained thus, again, its experimental study is a necessary step in the advancement of the field of polarimetric vacuum birefringence searches.

The phase sensitivities reached by differential-arm interferometers and by VMB polarimeters are within an order of magnitude for much of their measurement bands.  This is due to the significant effort and resources spent in the field of gravitational-wave detection to reduce displacement noise, and provides a techincal case to accompany an already intriguing science case\cite{Fouche2016} for alternate vacuum polarization search topologies. For polarimeter searches, our study suggests that reaching the necessary sensitivity poses a different set of challenges, firstly that of overcoming cavity-mirror birefringence noise.

\section{Acknowledgments}

This research has been partially supported by ANR (Grant No. ANR-14-CE32-0006) in the framework of the ``Programme des Investissements d'Avenir''.  We thank all the members of the BMV project and in particular M. Fouch\'e who greatly contributed in the recent years to the BMV experiment.  We also acknowledge the contributions of J. Renoud and M. Humbert.

\end{document}